\documentclass[twocolumn,aps,showpacs,nofootinbib,tightenlines]{revtex4}
\usepackage{amsmath}
\usepackage{amssymb}
\usepackage{graphicx}
\usepackage{subfigure}
\usepackage{mathrsfs}
\graphicspath{{figures/}}

\begin{document}
\title{Gravitational Contributions to the Running of Gauge Couplings}
\author{Yong Tang}
\author{Yue-Liang Wu}
\affiliation{ Kavli Institute for Theoretical Physics China (KITPC)
\\ Key Laboratory of Frontiers in Theoretical Physics, Institute of
Theoretical Physics, Chinese Academy of Science, Beijing, 100190,
P.R.China}
\begin{abstract}
Gravitational contributions to the running of gauge couplings are
calculated by using different regularization schemes. As the $\beta$
function concerns counter-terms of dimension four, only quadratic
divergences from the gravitational contributions need to be
investigated. A consistent result is obtained by using a
symmetry-preserving loop regularization with string-mode regulators
which can appropriately treat the quadratic divergences and
preserve non-abelian gauge symmetry. The harmonic gauge condition
for gravity is used in both diagrammatical and background field
calculations, the resulting gravitational corrections to the $\beta$
function are found to be nonzero, which is different from previous
results presented in the existing literatures.
\end{abstract}
 \maketitle
 {\bf Introduction}: Enclosing general relativity into the framework of quantum field
 theory is one of the most interesting and frustrating questions. Since
 its coupling constant $\kappa$ is of negative mass dimension,
 general relativity has isolated itself from the renormalizable
 theories. Renormalization is tightly connected with symmetry and
 divergency of loop diagrams.
 It was until the invention of dimensional regularization\cite{DR} that
 divergences of gravitation were not investigated systematically.
 Although, pure gravity at one-loop is
 free of physically relevant divergences \cite{Veltman},
 if higher-loop diagrams are considered, higher dimensional
 counter-terms are needed to add to the original lagrangian. With the
 hope that quantum gravity, coupled with other fields, might show
 some miraculous cancelations, Einstein-Scalar system was investigated
 by using dimensional regularization which makes all divergences
 reduced to logarithmic ones, and found to be non-renormalizable\cite{tHooft}.
 Later, Einstein-Maxwell, Einstein-Dirac, and Einstein-Yang-Mills
 systems were studied and shown to be also non-renormalizable\cite{van}.
 In spite of this theoretical inconsistence of general
 relativity and quantum field theory, we cannot
 deny that gravity has effect on ordinary fields and it contributes
 to the corrections of physical results since all matter has
 gravitational interaction. Treating general relativity as an effective field
 theory provides a practical way to look into quantum gravity's effects\cite{Donoghue}.

Recently, Robinson and Wilczek \cite{RW} calculated gravitational
corrections on gauge theory and arrived at an interesting
observation that at or beyond Planck scale, all gauge theories,
abelian or non-abelian, are asymptotically free due to gravitational
corrections to the running of gauge couplings. The result obtained
in ref.\cite{RW} has attracted one's attention. After reconsidering
gravitational effects on gauge theories, different conclusions were
reached by several groups. The result in\cite{RW} was derived in the
framework of background field method, as such a method has off-shell
and gauge-dependent problem, it was shown in \cite{Pietrykowski}
that the result obtained in \cite{RW} is gauge dependent, and the
gravitational correction to $\beta$ function at one-loop order is
absent in the harmonic gauge or general $\xi$-gauge. Using
Vilkovisky-DeWitt method, it was found in \cite{Toms} that the
gravitational corrections to the $\beta$ function also vanishes in
dimensional regularization. Late on, a diagrammatical calculation
for two and three point functions was performed in \cite{Ebert} to
obtain the $\beta$ function by using cut-off and dimensional
regularization schemes, as a consequence, the same conclusion was
yielded that quadratic divergences are absent.

In all these calculations, the crucial question concerns how to
appropriately treat quadratic divergences. If gravitational
corrections have no quadratic divergences, there would be no
change to the $\beta$ function of gauge couplings. Since different
regularization schemes treat quadratic divergences differently,
we would like apply in this paper the symmetry-preserving Loop
Regularization (LR) with string-mode regulators\cite{YLW} to
evaluate the gravitational contributions to the running of gauge
couplings. This is because the LR method has been shown to preserve
both the non-abelian gauge symmetry and the divergent behavior of
original integrals, it has be applied to consistently obtain all
one-loop renormalization constants of non-abelian gauge theory and
QCD $\beta$ function\cite{CW}, to derive the dynamically generated
spontaneous chiral symmetry breaking in chiral effective field
theory\cite{DW}, to clarify the ambiguities of quantum chiral
anomalous\cite{MW1} and the topological Lorentz/CPT violating
Chern-Simons term\cite{MW2}, and to verify the supersymmetric Ward
identities and non-renormalization theorem in supersymmetric field
theories\cite{CTW}.

In the following, after briefly introducing the symmetry-preserving
Loop Regularization (LR) method proposed in \cite{YLW}, and
presenting the general formalism needed for considering
gravitational effects, we first make a careful check to recover the
results obtained in \cite{Pietrykowski,Toms,Ebert} by using
dimensional and cut-off regularization schemes. Then by using the LR
method with both the diagrammatical and background field method
calculations, we present new results. Finally, we arrive at a
conclusion which differs from the ones presented in the existing
literatures.

{\bf Loop Regularization}: all loop calculations of Feynman diagrams
can be reduced, after Feynman parametrization and momentum
translation, into some simple scalar and tensor type loop integrals.
For divergent integrals, a regularization is needed to make them
physically meaningful. There are many kinds of regularization
schemes in literature. For the reason mentioned above, we shall
adopt the Loop Regularization method. In the LR method, the crucial
concept is the introduction of the irreducible loop
integrals(ILIs)\cite{YLW} which are defined, for example, at
one-loop level as
\begin{eqnarray}
I_{-2\alpha}(\mathcal{M}^{2})&=&\int d^4k\frac{1}{(k^2-\mathcal{M}^{2})^{2+\alpha}},\nonumber\\
I_{-2\alpha\ \mu\nu}(\mathcal{M}^{2})&=&\int
d^4k\frac{k_{\mu}k_\nu}{(k^2-\mathcal{M}^{2})^{3+\alpha}}
\end{eqnarray}
and higher rank of tensor integrals, with $\alpha=-1,0,1,...$. Here
$I_2$ and $I_0$ denote the quadratic and logarithmic divergent ILIs
respectively. In general, the divergent integrals are meaningless.
To see that, let us examine tensor and scale type quadratically
divergent ILIs $I_{2\mu\nu}(0)$ and $I_2(0)$, one can always write
down, from the Lorentz decomposition, the general relation that
$I_{2\mu\nu}(0) = a g_{\mu\nu} I_2(0)$ with $a$ to be determined
appropriately. When naively multiplying $g^{\mu\nu}$ on both sides
of the relation, one yields $a =1/4$, which is actually no longer
valid due to divergent integrals. To demonstrate that, considering
the zero component $I_{2 00}$ of tensor ILIs $I_{2\mu\nu}$, and
performing an integration over the momentum $k^0$ for both $I_{2
00}$ and $I_2$, it is then not difficult to found after comparing
both sides of the relation that $a=1/2$ which should hold as the
integrations over $k^0$ for $I_{2 00}$ and $I_2$ are convergent. To
check its consistence, considering the vacuum polarization of QED.
In terms of the ILIs, the vacuum polarization is given by
\begin{eqnarray}\label{VaP}
\Pi_{\mu\nu}&=&-4e^2\int
dx\big{[}2I_{2\mu\nu}(m)-I_2(m)g_{\mu\nu}\nonumber\\
&+&2x(1-x)(p^2g_{\mu\nu}-p_{\mu}p_{\nu})I_{0}(m)\big{]}
\end{eqnarray}
which shows that only quadratic divergences violate gauge
invariance.

Note that the cut-off regularization together with the imposition of
the Ward-Takahashi identities could also give the gauge invariant
result of vacuum polarization in QED by introducing a photon mass
renormalization as shown in\cite{JR}, but the cut-off
regularization, itself, cannot respect Ward-Takahashi identities
which are crucial for unitarity of S-matrix. Therefore, the cut-off
regularization alone violates gauge invariance.

However, if an explicit regularization has a property that the
regularized quadratic divergences satisfy the consistency
condition\cite{YLW}
\begin{eqnarray}
I^{R}_{2\mu\nu}=\frac{1}{2}g_{\mu\nu}I^{R}_{2}
\end{eqnarray}
namely $a=1/2$, then the vacuum polarization becomes gauge
invariant. Here the superscript "R" denotes the regularized ILIs.
Though dimensional regularization preserves gauge invariance with
the above consistency condition and can handle with quadratic
divergences, nevertheless, it is not originally intended to maintain
the divergent behavior of original integrals. The LR method has been
shown to satisfy the consistency conditions and meanwhile maintain
the divergent behavior of original integrals\cite{YLW}. The
regularized divergent ILIs in loop regularization have the following
explicit results\cite{YLW}
\begin{eqnarray}\label{sILI}
& & I_{2\mu\nu}^R = \frac{1}{2} g_{\mu\nu} I_2^R,\qquad I_{0\mu\nu}
= \frac{1}{4} g_{\mu\nu}I_0^R \nonumber \\
& & I_2^R = \frac{-i}{16\pi^2}
\bigg{[}M_c^2-\mu^2[\ln{\frac{M_c^2}{\mu^2}}
-\gamma_w+1+y_2(\frac{\mu^2}{M_c^2})]\bigg{]}\nonumber \\
& & I_0^R = \frac{i}{16\pi^2}
\bigg{[}\ln{\frac{M_c^2}{\mu^2}}-\gamma_w+y_0(\frac{\mu^2}{M_c^2})\bigg{]}
\end{eqnarray}

with $\mu^2=\mu_s^2+\mathcal{M}^2$,
$\gamma_w=\gamma_E=0.5772\cdots$, and
\begin{eqnarray}
& & y_0(x)=\int_0^x d\sigma \frac{1-e^{-\sigma}}{\sigma},
 \quad  y_1(x)=\frac{e^{-x}-1+x}{x}\nonumber \\
& & y_2(x)=y_0(x)-y_1(x),\quad \lim_{x\rightarrow0}y_i(x)\rightarrow
0
\end{eqnarray}
Here the scales $M_c$ and $\mu_s$ play the rule of characteristic
energy scale and sliding energy scale. A detailed derivation is
referred to ref.\cite{YLW}.

It is interesting to notice that to understand the above consistency
condition for the quadratic divergences in dimensional
regularization, one may simply base on the observation that the
quadratic divergences for $D=4$ show up as logarithmic divergences
for $D=2$, thus the quadratic divergences may be dealt with by
naively taking $D=2$ as discussed in\cite{MV} and further developed
in\cite{JJ}. Nevertheless, such a treatment on the quadratic
divergences in dimensional regularization can only be regarded as a
special alternative and incidental explanation on the relation
eq.(3) rather than a systematical approach.


{\bf General Formalism}: The action of Einstein-Yang-Mills theory is
\begin{equation}\label{LEYM}
\textrm{S}=\int
d^{4}x\sqrt{-\textbf{g}}\left[\frac{1}{\kappa^2}\textbf{R}
-\frac{1}{4}\textbf{g}^{\mu\alpha}\textbf{g}^{\nu\beta}\mathcal{F}^{a}_{\mu\nu}\mathcal{F}^{a}_{\alpha\beta}\right]
\end{equation}
where \textbf{R} is Ricci scalar and $\mathcal{F}^{a}_{\mu\nu}$ is
the Yang-Mills fields strength
$\mathcal{F}_{\mu\nu}=\nabla_{\mu}\mathcal{A}_{\nu}-\nabla_{\nu}\mathcal{A}_{\mu}
-ig[\mathcal{A}_{\mu},\mathcal{A}_{\nu}]$. It is hard to quantize
this lagrangian because of gravity-part's non-linearity and
minus-dimension coupling constant $\kappa=\sqrt{16\pi\textrm{G}}$.
Usually, one expands the metric tensor around a background metric
$\bar{g}_{\mu\nu}$ and treats graviton field as quantum fluctuation
$h_{\mu\nu}$ propagating on the background space-time determined by
$\bar{g}_{\mu\nu}$.
\begin{equation}
\textbf{g}_{\mu\nu}=\bar{g}_{\mu\nu}+\kappa h_{\mu\nu}
\end{equation}
The above expansion is exact, but the expansions of inverse metric
and determinant are approximate with ignoring higher-order terms. To
the second order in $\kappa$,
\begin{eqnarray}
\textbf{g}^{\mu\nu}&=&\bar{g}^{\mu\nu}-\kappa h^{\mu\nu}+\kappa ^2
h^{\mu}_{\alpha}h^{\alpha\nu}+...\\
\sqrt{-\textbf{g}}&=&\sqrt{-\bar{g}}[1+\frac{1}{2}\kappa
h-\frac{1}{4}\kappa ^2 (h^{\mu\nu}h_{\mu\nu}-\frac{1}{2}h^2)...]
\end{eqnarray}
In fact, the above expansion is an infinite series and the
truncation is up to the question considered. These infinite series
partly indicate that gravity is not renormalizable. From this
perspective, it is more accurate to say that the system is treated
as an effective field theory. After assembling the same order terms
in $\kappa$, we could get the lagrangian for usual quantization.

{\bf Diagrammatical Calculation}: Let us set $\bar{g}_{\mu\nu}=\eta
_{\mu\nu}$, where $\eta _{\mu\nu}$ is the Minkowski metric.
$h_{\mu\nu}$ is interpreted as graviton field, fluctuating in flat
space-time. The lagrangian can be arranged to different orders of
$h_{\mu\nu}$ or $\kappa$. The free part of gravitation is of order
unit and gives the graviton propagator\cite{Veltman}
\begin{equation}
P_{G}^{\mu\nu\rho\sigma}(k)=\frac{i}{k^2}[g^{\nu\rho}g^{\mu\sigma}+g^{\mu\rho}g^{\nu\sigma}
-g^{\mu\nu}g^{\rho\sigma}]
\end{equation}
in the de Donder harmonic gauge
\[
C^{\mu}=\partial_{\nu}h^{\mu\nu}-\frac{1}{2}\partial^{\mu}h^{\nu}_{\nu}=0
\]
For simplicity, in the following, the metric $g^{\mu\nu}$ is
understood as $\eta^{\mu\nu}$. The interactions of gauge field and
gravity field are determined by expanding the second term of the
lagrangian (\ref{LEYM}). And various vertex could be derived.
Details of these Feynman rules are referred to \cite{Ebert}. Using
the traditional Feynman diagram calculations, we can compute the
$\beta$ function by evaluating two and three point functions of
gauge fields. These green functions are general divergent, so
counter-terms are needed to cancel these divergences. The relevant
counter-terms to the $\beta$ function are
\begin{eqnarray}
& & T^{\mu\nu} = i\delta_{ab}Q^{\mu\nu} \delta_{2},\quad
T^{\mu\nu\rho}
= gf^{abc}V_{qkp}^{\mu\nu\rho} \delta_{1} \nonumber \\
& & Q^{\mu\nu} \equiv q^{\mu}q^{\nu}-q^2g^{\mu\nu} \\
& & V_{qkp}^{\mu\nu\rho} \equiv
g^{\nu\rho}(q-k)^{\mu}+g^{\rho\mu}(k-p)^{\nu}
+g^{\mu\nu}(p-q)^{\rho} \nonumber
\end{eqnarray}
The $\beta$ function is defined as
\[\beta(g)=g\
\mu\frac{\partial}{\partial\mu}(\frac{3}{2}\delta_{2}-\delta_{1})\]
In gauge theories without gravity, the counter-terms are
logarithmically divergent as the quadratic divergences cancel each
other due to the gauge symmetry. However, if gravitational
corrections are taken into account, divergent behavior becomes
different. On dimensional ground, it is known that quadratic
divergences can appear and will contribute to the counter-terms
defined above, so that they will also lead to the corrections to the
$\beta$ function. Although gravitational corrections contain
logarithmic divergences, these divergences are multiplied by
high order momentum. So logarithmic divergences will lead us to
introduce counter-terms of six dimension, like
$Tr[(D_{\mu}F_{\nu\rho})^{2}]$, $Tr[(D^{\mu}F_{\mu\nu})^{2}]$ and
$Tr[F_{\mu}^{ \nu}F_{\nu}^{ \rho}F_{\rho}^{ \mu}]$. These terms do
not affect $\beta$ function of gauge coupling constant. So, in later
calculations, we will omit the logarithmic divergences and only pay
our attention to the quadratic divergences. For convenience, the
gravitational contributions are labeled with a superscript $\kappa$.
\begin{equation}\label{grbeta}
\Delta\beta^{\kappa}=g\
\mu\frac{\partial}{\partial\mu}(\frac{3}{2}\delta^{\kappa}_{2}-\delta^{\kappa}_{1})
\end{equation}

It can be shown that at one-loop level, gravity will contribute to
two and three point functions from five diagrams (see Fig.~1).
\begin{figure}[t]
    \subfigure[]
              {\label{TwoPa}\includegraphics[scale=0.7]{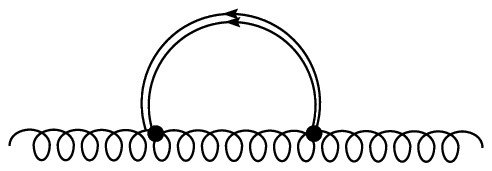}}
    \subfigure[]
              {\label{TwoPb}\includegraphics[scale=0.7]{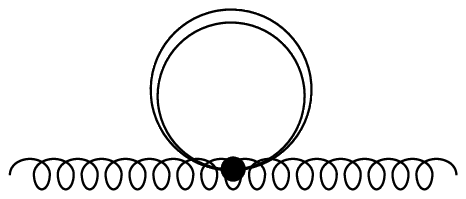}}\\
    \subfigure[]
              {\label{ThreePa}\includegraphics[scale=0.62]{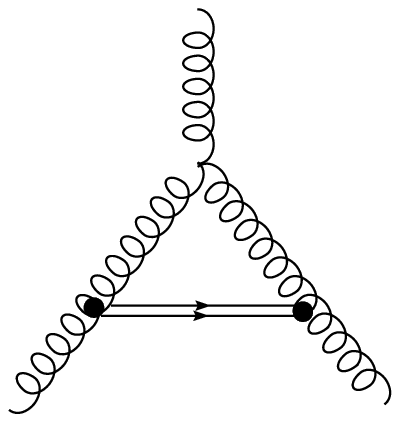}}
    \subfigure[]
              {\label{ThreePb}\includegraphics[scale=0.62]{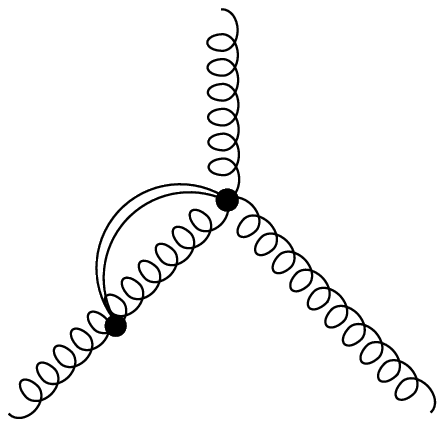}}
    \subfigure[]
              {\label{ThreePc}\includegraphics[scale=0.62]{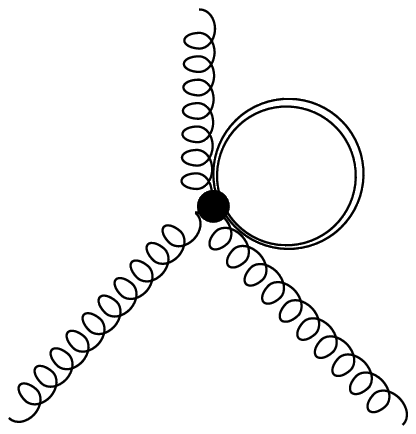}}
 \caption{Feynman diagrams that gravitation contributes at one loop order
  to gauge two and three point functions.}
\end{figure}
The first two diagrams are corresponding to two point function, and
the other three diagrams are for three point function. One can
easily show that Fig.~\ref{ThreePa} has no quadratic divergences
and only logarithmic divergences are left. For Fig $\ref{ThreePb}$,
as one of the end of graviton propagator could be attached to any
gauge external leg, there are two similar contributions which can be
obtained with a cycle for the momenta and their corresponding
Lorentz index. In harmonic gauge, the two point functions from
Fig.~1(a) and Fig.~1(b) are found in terms of ILIs to be
\begin{eqnarray}\label{ILI2a}
& & T^{(a)\mu\nu} = 2\kappa^2\int dx\biggl{[}Q^{\mu\nu}\left[I_{2}
+q^2(3x^2-x)I_{0}\right]\nonumber\\
&& + q^{\mu}q_{\rho}I^{\nu\rho}_{2}+q^{\nu}q_{\rho}I^{\mu\rho}_{2}
-g^{\mu\nu}q_{\rho}q_{\sigma}I^{\rho\sigma}_{2}-q^2I^{\mu\nu}_{2}\bigg{]}(\mathcal{M}^{2}_q)\nonumber\\
& & T^{(b)\mu\nu} = -3\kappa^2 Q^{\mu\nu} I_{2}(0)
\end{eqnarray}
and the three point functions from Fig.~$\ref{ThreePb}$ and Fig.~$\ref{ThreePc}$ are found, when keeping only the quadratically divergent terms, to be
\begin{eqnarray}
& &T^{(d)\mu\nu\rho} = ig\kappa^2 \bigg{\{} -V_{qkp}^{\mu\nu\rho}
I_{2}(0)
+ \nonumber \\
& & \int
dx\biggl{[}\biggl{(}g^{\mu\nu}q_{\sigma}I^{\rho\sigma}_{2}
-g^{\nu\rho}q_{\sigma}I^{\mu\sigma}_{2}+q^{\rho}I^{\mu\nu}_{2}
-q^{\mu}I^{\nu\rho}_{2}\biggl{)}(\mathcal{M}_{q}^{2})\nonumber \\
& & +\biggl{(}g^{\nu\rho}k_{\sigma}I^{\mu\sigma}_{2}
-g^{\rho\mu}k_{\sigma}I^{\nu\sigma}_{2}+k^{\mu}I^{\nu\rho}_{2}
-k^{\nu}I^{\rho\mu}_{2}\biggl{)}(\mathcal{M}_{k}^{2})\nonumber\\
& &+ \biggl{(}g^{\rho\mu}p_{\sigma}I^{\nu\sigma}_{2}
-g^{\mu\nu}p_{\sigma}I^{\rho\sigma}_{2} +p^{\nu}I^{\rho\mu}_{2}
-p^{\rho}I^{\mu\nu}_{2}\biggl{)}(\mathcal{M}_{p}^{2})\biggl{]}
\bigg{\}} \nonumber \\
& & T^{(e)\mu\nu\rho} = 3ig\kappa^2 V_{qkp}^{\mu\nu\rho} I_{2}(0)
\end{eqnarray}
with $\mathcal{M}^{2}_q=x(x-1)q^{2}$. Contraction is performed by
using FeynCalc package\cite{feyncalc}.

Now we shall apply the different regularization schemes to the
divergent ILIs. In cut-off regularization, when keeping only
quadratically divergent terms, one has
\begin{equation}
I_{2}^{R \mu\nu} = \frac{1}{4}g^{\mu\nu} I_2^R,\quad I_{2}^{R}
\simeq \frac{i}{16\pi^{2}} g^{\mu\nu}\Lambda^{2}
\end{equation}
The resulting two and three point functions are
\begin{eqnarray}
& & T^{(a+b)\mu\nu}_{cutoff} \equiv
T^{(a)\mu\nu}_{cutoff}+T^{(b)\mu\nu}_{cutoff} \approx 2Q^{\mu\nu}
\kappa^2\int dx \nonumber \\
& & \biggl{[}\frac{1}{2}\frac{i}{16\pi^{2}}
\Lambda^{2}+\left[\frac{i}{16\pi^{2}}\Lambda^{2}-\frac{3}{2}\frac{i}{16\pi^{2}}\Lambda^{2}\right]\biggl{]}=0
\\
& & T^{(d+e)\mu\nu\rho}_{cutoff} \equiv T^{(d)\mu\nu\rho}_{cutoff} +
T^{(e)\mu\nu\rho}_{cutoff} \approx 0
\end{eqnarray}
which agrees with the result obtained in \cite{Ebert}.

In dimensional regularization, where $I_{2}^R(0)=0$ and
$I_{2\mu\nu}^R=\frac{1}{2}g_{\mu\nu}I_{2}^R$, the two and three
point functions are found to be
\begin{eqnarray}
& & T^{(a+b)\mu\nu}_{DR} \approx 4\kappa^2 Q^{\mu\nu} \int
dxI_{2}^R(\mathcal{M}^{2}_q),\\
& & T^{(d+e)\mu\nu\rho}_{DR} =2ig\kappa^2\int dx
\biggl{[}(g^{\mu\nu}q^{\rho}-q^{\mu}g^{\nu\rho})I_{2}^R(\mathcal{M}_{q}^{2}) \\
& &
+(g^{\nu\rho}k^{\mu}-k^{\nu}g^{\rho\mu})I_{2}^R(\mathcal{M}_{k}^{2})+
(g^{\rho\mu}p^{\nu}-p^{\rho}g^{\mu\nu})I_{2}^R(\mathcal{M}_{p}^{2})\biggl{]}
\nonumber
\end{eqnarray}
where the regularized quadratic divergence in dimensional
regularization behaves as the logarithmic one
\begin{eqnarray}
& & I_{2}^R(\mathcal{M}^{2}_q)|_{DR} = - \frac{-i}{16\pi^2}
\mathcal{M}^{2}_q [\frac{2}{\varepsilon} -\gamma_E+1+
O(\varepsilon)]
\end{eqnarray}

So far, we have shown that in both cut-off regularization and
dimensional regularization, there are no gravitational corrections
to the gauge $\beta$ function, which agrees with the ones yielded in
\cite{Pietrykowski,Toms,Ebert}.

We now make a calculation by using loop regularization. With the
consistency condition $I_{2\mu\nu}^R=\frac{1}{2}g_{\mu\nu}I_{2}^R$,
we obtain for two and three point functions
\begin{eqnarray}
& & T^{(a+b)\mu\nu}_{LR}= 2\kappa^2 Q^{\mu\nu} \int
dx\bigg{[}-\frac{3}{2}I^{R}_{2}(0) + 2I^{R}_{2}(\mathcal{M}^{2}_q) \nonumber \\
&&
\qquad +q^2(3x^2-x)I^{R}_{0}(\mathcal{M}^{2}_q)\bigg{]} \\
& & T^{(d+e)\mu\nu\rho}_{LR} =2ig\kappa^2\int dx\biggl{[}
\frac{1}{2} V_{qkp}^{\mu\nu\rho} I^{R}_{2}(0) \nonumber
\\
& & +
(g^{\mu\nu}q^{\rho}-q^{\mu}g^{\nu\rho})I^R_{2}(\mathcal{M}_{q}^{2})
+(g^{\nu\rho}k^{\mu}-k^{\nu}g^{\rho\mu})I^R_{2}(\mathcal{M}_{k}^{2})
\nonumber \\
&& +
(g^{\rho\mu}p^{\nu}-p^{\rho}g^{\mu\nu})I^R_{2}(\mathcal{M}_{p}^{2})\biggl{]}
\end{eqnarray}
from which we can directly read off the two-point and three-point
counter-terms $\delta^{\kappa}_{2}$ and $\delta^{\kappa}_{1}$
respectively
\begin{eqnarray}
& &
\delta^{\kappa}_{2}=\kappa^{2}\frac{1}{16\pi^2}\bigg{[}M_c^2-\mu^2[\ln\frac{M_c^2}{\mu^2}-\gamma_w+1 +
y_2(\frac{\mu^2}{M_c^2})]\bigg{]} \nonumber \\
& &
\delta^{\kappa}_{1}=\kappa^{2}\frac{1}{16\pi^2}\bigg{[}M_c^2-\mu^2[\ln\frac{M_c^2}{\mu^2}-\gamma_w+1
+ y_2(\frac{\mu^2}{M_c^2}) ]\bigg{]} \nonumber \\
\end{eqnarray}
Putting the leading quadratically divergent part of $\delta^{\kappa}_{1}$ and $\delta^{\kappa}_{2}$ into
Eq.($\ref{grbeta}$), we obtain the gravitational corrections to the
gauge $\beta$ function
\begin{eqnarray}\label{betacrrtn}
\Delta\beta^{\kappa}=-g\kappa^2\frac{\mu_s^2}{16\pi^2}
\end{eqnarray}
which shows that there are gravitational quadratic corrections to
the gauge $\beta$ function when loop regularization method is
adopted to evaluate the quadratic divergent integrals, which is
different form the results yielded by using the cut-off and
dimensional regularization schemes.

{\bf Background Field Method}: we have also performed a calculation
by using the background field formalism. By using the loop
regularization method and taking the harmonic gauge
condition, we then obtain the same $\beta$ function correction as
eq.(\ref{betacrrtn}).

In general, the total $\beta$ function of gauge field theories
including the gravitational effects may be written as follows
\begin{equation}
\beta^\kappa = \mu\frac{dg}{d\mu} = - \frac{b_0 }{16\pi^2}g^3 -\frac{\mu^2}{16\pi^2}g\kappa^2
\end{equation}
The interesting feature of gauge theory interactions is the possible
gauge couplings unification\cite{U1,U2} at ultra-high energy scale
when the gravitational effects are absent. To illustrate the effects
of gravitational contributions, we plot in Fig.~2 the running of
gauge couplings based on the minimal supersymmetric standard model
(MSSM) in which the three gauge couplings are unified at the energy
scale $10^{16}$ GeV.

\begin{figure}
\includegraphics[scale=1.0]{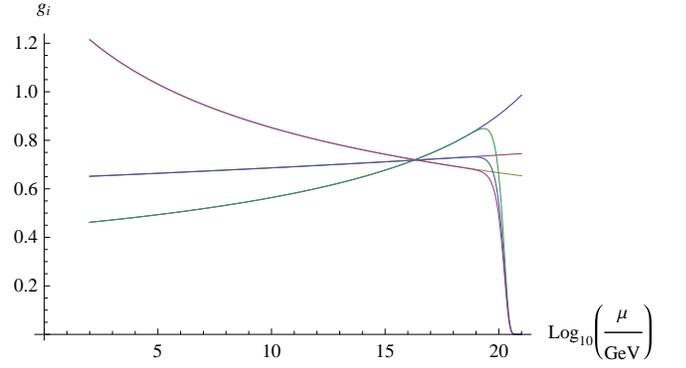}
 \caption{An illustration of gravitational contributions to the running of gauge couplings in the MSSM}
\end{figure}

Where the running of gauge couplings in the MSSM without gravitational
contributions is known to be ($\alpha_{i}=\frac{g^2_i}{4\pi}$)
\begin{eqnarray}
\alpha^{-1}_{e}(\mu)&=&\alpha^{-1}_{e}(M)+\frac{33}{10\pi}\ln{\frac{M}{\mu}}\nonumber \\
\alpha^{-1}_{w}(\mu)&=&\alpha^{-1}_{w}(M)+\frac{1}{2\pi}\ln{\frac{M}{\mu}}\\
\alpha^{-1}_{s}(\mu)&=&\alpha^{-1}_{s}(M)-\frac{3}{2\pi}\ln{\frac{M}{\mu}}
\nonumber
\end{eqnarray}
with experimental input at $M_Z$
\begin{eqnarray}
&&\alpha^{-1}_{e}(M_{Z})=58.97\pm0.05, \nonumber \\
&&\alpha^{-1}_{w}(M_{Z})=29.61\pm0.05,\\
&&\alpha^{-1}_{s}(M_{Z})=8.47\pm0.22 \nonumber
\end{eqnarray}
It is noticed from Fig.~2 that only in the energy scale near and
above Planck energy scale ($\sim 10^{19}\textrm{GeV}$) the gravitational
effects become significant.

{\bf Conclusions}: we have investigated the gravitational
contributions to the running of gauge couplings by adopting
different regularization schemes. From the above explicit
calculations, it is not difficult to see that the different
conclusions resulted from different regularization schemes mainly
arise from the treatment for the quadratic divergent integrals.
Since LR takes the consistency condition for the regularized
quadratic divergent ILIs that $I_{2\mu\nu}^R =
\frac{1}{2}g_{\mu\nu}I_2^R$, and maintains the quadratically divergent
behavior for the regularized ILIs $I_2^R$, it then gives a nonzero
gravitational quadratic corrections to the gauge $\beta$ function
with the de Donder harmonic gauge condition. If there is an other
regularization scheme which have the same properties as above, it
will give the same result. Also it is of interest to note that the
quadratic corrections in pure gauge theory are know to cancel each
other due to the consistency conditions of gauge symmetry\cite{YLW},
while the gravitational quadratic corrections to the running of
gauge couplings are all gauge invariant and there is no symmetry to
forbid their existence. In the de Donder harmonic
gauge, ghost fields for the gravity are not coupled with gauge fields at one-loop order, which simplifies the calculation. We then come to the conclusion that at one-loop order in de Donder harmonic all gauge theories at or beyond Planck scale become asymptotically free as the resulting $\beta$ function from the
gravitational corrections are negative in this case. Note that the result may have gauge condition dependence, we \cite{YongWu} are also looking in gauge condition independent formalism \cite{Vilkovisky,DeWitt}

\acknowledgments This work was supported in part by the National
Science Foundation of China (NSFC) under the grant \# 10821504,
10491306, 10975170 and the Project of Knowledge Innovation Program (PKIP) of
Chinese Academy of Science.


\begin{thebibliography}{99}
 \bibitem{DR}G.'t Hooft and M.~Veltman, Nucl.Phys. \textbf{B}44 189 (1972).
 \bibitem{Veltman}G.'t Hooft and M.~Veltman, Ann. Poincare Phys. Theor. \textbf{A}20, 69
                  (1974); M. Veltman, Lecture Notes of Method in Field Theory, Les
                  Houches, 1975.
 \bibitem{tHooft}G.'t Hooft, Nucl. Phys. \textbf{B}62, 444 (1973).
 \bibitem{van}S.~Deser and P.van Nieuwenhuizen, Phys. Rev. \textbf{D}10, 401 (1974);
              S.~Deser, H.S.~Tsao and P.van Nieuwenhuizen, Phys. Rev. \textbf{D}10, 3337 (1974).
 \bibitem{Donoghue}J.~F.~Donoghue, Phys.
  Rev. Lett. {\bf 72} 2996 (1994); Phys. Rev. \textbf{D}50, 3874 (1994).
 \bibitem{RW}S.~P.~Robinson and F.~Wilczek, Phys.
  Rev. Lett. {\bf 96}, 231601 (2006).
 \bibitem{Pietrykowski}A.~R.~Pietrykowski,Phys. Rev. Lett. {\bf 98}, 061801 (2007).
 \bibitem{Toms}D.J.~Toms, Phys. Rev. \textbf{D}76, 045015 (2007).
 \bibitem{Ebert}D.~Ebert, J.~Plefka and A.~Rodigast, Phys. Lett. \textbf{B}660, 579
 (2008).
 \bibitem{YLW}Y.~L.~Wu, Int.J. Mod. Phys. \textbf{A}18, 5363 (2003), [arXiv:hep-th/0209021]; \\
             Y.~L.~Wu, Mod. Phys. Lett. \textbf{A}19, 2191 (2004), [arXiv:hep-th/0311082].
 \bibitem{JR}J.~M.~Jauch and F.~Rohrlich, ''The Theory of Photons and
 Electrons", Springer-Verlag, New York, Heidelberg, Berlin, 1976.
 \bibitem{MV} M.~Veltman, Acta Physica Polonica {\bf B12}, 437 (1981).
 \bibitem{JJ} I.~Jack and D.R.T.~Jones, Nucl. Phys. {\bf B342}, 127 (1990).
 \bibitem{CW} J.~W.~Cui and Y.~L.~Wu,  Int. J. Mod. Phys. A {\bf 23}, 2861 (2008)  [arXiv:0801.2199 [hep-ph]].
 \bibitem{DW} Y.~B.~Dai and Y.~L.~Wu, Eur. Phys. J. {\bf C39}, 1 (2004).
 \bibitem{MW1} Y.~L.~Ma and Y.~L.~Wu, Int. J. Mod. Phys. {\bf A21}, 6383 (2006), [arXiv:hep-ph/0509083].
 \bibitem{MW2} Y.~L.~Ma and Y.~L.~Wu,  Phys. Lett. {\bf B647}, 427(2007), [arXiv:hep-ph/0611199].
 \bibitem{CTW} J.~W.~Cui, Y.~Tang and Y.~L.~Wu, Phys. Rev. \textbf{D} 79, 125008 (2009).
 \bibitem{feyncalc}R.~Mertig, M.~B\"ohm, and A.~Denner, Comput. Phys. Commun. {\bf 64}, 345(1991).
 \bibitem{U1} H.~Georgi, H.Quinn and S.~Weinberg, Phys. Rev. Lett. {\bf
 33}, 451 (1974).
 \bibitem{U2} S.~Dimopoulos, S.~Raby, and F.~Wilczek, Phys. Rev. \textbf{D} 24, 1681 (1981).
 \bibitem{YongWu} Yong Tang, Yue-Liang Wu, in preparation.
 \bibitem{Vilkovisky}
  G.~A. Vilkovisky, Nucl. Phys. B {\bf 234}, 125 (1984),  {\it The Quantum Theory of Gravity\/}, edited by S.~M. Christensen (Adam Hilger, Bristol, 1984).
 \bibitem{DeWitt}
  B.~S. DeWitt, in {\it Quantum Field Theory and Quantum Statistics, Volume 1\/}, edited by I.~A. Batalin, C.~J. Isham, and G.~A. Vilkovisky (Adam Hilger, Bristol, 1987).

\end{thebibliography}
\end{document}